# ADAPTIVE OPTICS CORRECTION OF THE WAVEFRONT DISTORTIONS INDUCED BY SEGMENTS MISALIGNMENT IN EXTREMELY LARGE TELESCOPE

Natalia Yaitskova[*] and Christophe Verinaud[**]

European Southern Observatory
Karl Schwarzschildstr. 2 D-85748 Garching bei Muenchen, Germany

**ABSTRACT**

The capability of the adaptive optics to correct for the segmentation error is analyzed in terms of the residual wavefront RMS and the power spectral density of the phase. The analytical model and the end-to-end simulation give qualitatively equal results justifying the significance of the geometrical matching between segmentation geometry and the actuators/subaperture distribution of the adaptive optics. We also show that the design of the wavefront sensor is rather critical.

## 1. INTRODUCTION

The wavefront control of Extremely Large Telescope (ELT) includes three main units: adaptive optics (AO), active optics and phasing camera. Each unit is meant to correct for the different components of the distorted wavefront: AO is responsible for an atmospheric turbulence, active optics – for the misalignment of the telescope mirrors and their deformation, and the phasing camera – for the misalignment of the individual segments in the segmented mirrors. The corresponding control loops run independently. Disentangling of the components is based on their difference in temporal and spatial bandwidths. Nevertheless, each wavefront control unit is affected to some extend by the total wavefront, and the alien components are usually considered as the external noise for a given control unit. On the other hand, a situation is possible when the control units are "helping" each other. Thus, the partial correction of the telescope aberrations can be performed moving the segments of a primary mirror, i.e. by a phasing unit. The WaveFront Sensor (WFS) of AO unit will see the telescope aberrations and segments misalignments; hence the AO deformable mirror (DM) will try to compensate for them.

The goal of the present study is to determine the capability of the adaptive optics to correct for the segmentation piston errors. Our objectives are the shape of the residual wavefront, RMS and power spectral density (PSD) of the residual phase. In a second section we estimate the limits to which the continuous DM can reconstruct the step-like wavefront. In this part we abstract ourselves from the wavefront sensing module and assume that the wavefront is known exactly. The DM response is modeled by an influence function with the shape of a cubic spline. This part is done in pure analytical approach.

In a third section we present the result of simulations including AO WFS. We consider two types of WFS: Shack-Hartmann and Pyramid. The AO DM is modeled analogous to one in the second section. The goal here is to validate an adequacy of the analytical model and to compare two types of the sensors.

This work is particularly interesting for high contrast imaging applications, like the search for extra-solar planets, when any systematic error in the final image can be deleterious to the achievable contrast. We emphasize this in a section 3.3 while comparing the performance of different AO WFS in terms of a residual PSF halo.

---

Email: [*]nyaitsko@eso.org and [**]cverinau@eso.org

## 2. ANALYTICAL MODEL

### 2.1 Description of the model

Consider a segmented telescope where a segment numbered by index $j$ is described by a transmission function, $\theta_j(x - r_j)$, which takes on a unit value within the segment and zero value outside. In the case of segment piston errors the phase can be represented as a sum

$$\varphi(x - r_j) = \sum_{j=1}^{N} \delta_j \theta_j(x - r_j). \qquad (1)$$

Here $r_j$ describes the central position of segment with index $j$. Coefficient $\delta_j$ describes a piston error. We assume that the piston values on two different segments are statistically independent and identically distributed with a zero mean and standard deviation $\sigma$ :

$$\langle \delta_j \delta_i \rangle = \begin{cases} 0, & i \neq j \\ \sigma^2, & i = j \end{cases}. \qquad (2)$$

If the phasing loop is on the last expression is an approximation only.

For the DM we impose the following behavior:
(1) The actuator centered at the point $x_n$ is pushed to the value $-\varphi(x_n)$, where $\varphi(x_n)$ is the phase error at the point $x_n$ produced by the shift of the segment containing this point
(2) The shape of the DM created by pushing one separate actuator is described by a DM influence function $IF(x - x_n)$
(3) There is no actuator centered between segments

These assumptions allows us writing the residual after adaptive correction wavefront as

$$\varphi_{res}(x) = \sum_{j=1}^{N} \delta_j \theta_j(x - r_j) - \sum_n IF(x - x_n)\varphi(x_n) = \sum_{j=1}^{N} \delta_j \left[ \theta_j(x - r_j) - \sum_{\substack{m \\ x_n \in \theta_j}} IF(x - x_n) \right]. \qquad (3)$$

In a second item we performed first a sum over actuators belonging to one segment, and then over the segments.

Our goal is to define the upper limit of the DM performance. Therefore we use the cubic spline functions to model the DM behavior. Within the bounds of this model the DM can reconstruct low order aberration without artifact pinning errors, which appear if other shape of the influence function is used.[1] We use the cubic spline functions on the base $[-2d_a, 2d_a]$, which in one dimensional case are given by expression:[2]

$$IF(x) = \frac{1}{(2c+1)} \begin{cases} 1 + (4c - 2.5)x^2 + (-3c + 1.5)|x|^3 & 0 < x < d_a \\ (2c - 0.5)(2 - |x|)^2 + (-c + 0.5)(2 - |x|)^3 & d_a < x < 2d_a \\ 0 & x > 2d_a \end{cases} \qquad (4)$$

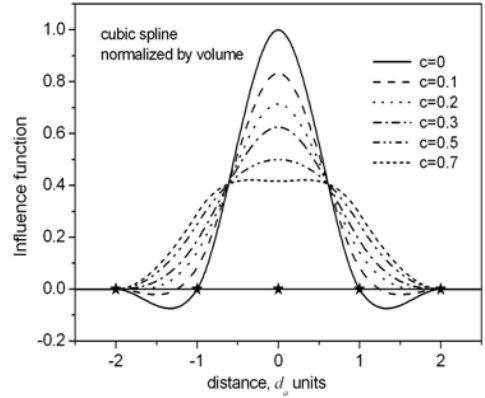

**Figure 1.** Deformable mirror influence function for different values of coupling coefficient. Stars position the central and neighboring actuators.

where $c \in Re$ is an inter-actuator coupling coefficient (shape parameter), $d_a$ is actuators' spacing. Note that to reconstruct the flat surface according to the algorithm given by Eq (3) the function $IF(x)$ must have a unit volume. The one dimensional cut of $IF(x)$ is shown in Figure 1.

### 2.2 Segment influence function and the shape of the residual wavefront

Let us introduce a new function which is a sum of the DM influence functions over the segment area. Let center it in the center of the segment:

$$IFs_j(\boldsymbol{x} - \boldsymbol{r}_j) \equiv \sum_{m,\, \boldsymbol{x}_m \in \theta_j} IF(\boldsymbol{x} - \boldsymbol{x}_m). \tag{5}$$

In the following we refer to it is as a *segment influence function*. The segment influence function depends on the density and the placement of the actuators, as well as on the shape of the segment and *IF* itself. The residual wavefront can be written as

$$\varphi_{res}(\boldsymbol{x}) = \sum_{j=1}^{N} \delta_j \left[ \theta_j(\boldsymbol{x} - \boldsymbol{r}_j) - IFs_j(\boldsymbol{x} - \boldsymbol{x}_j) \right]. \tag{6}$$

The simplest geometry for which function $IFs(\boldsymbol{x})$ is identical for all segments is a case of square segments and regular squared distribution of the actuators. Besides, the segment size, $d$, must be multiple of the actuator separation $d_a$: $d/d_a = m$, $m = 1, 2, 3...$ For odd $m$ the central actuator coincides with the center of a segment; for even $m$ the actuator pattern is shifted in both directions by $d_a/2$. With an increase of $m$ the segment influence function fits better the segment (Figure 2), and the DM reconstructs the segmentation error more precisely.

The corrected wavefront is shown in Figure 3. It is composed of oscillations, which are anti-symmetrical with respect to the intersegment border. Amplitude of the oscillations equals to the difference in piston errors between two corresponding segments, and the full width equals to $3d_a$. The shape of the oscillation depends on the coupling factor $c$. For $m>3$ phase in the middle of the segment is zero, due to the properties of cubic spline influence function. Similar shape is observed in a focal filtering approach.[3]

### 2.3 Residual RMS

From Eq. (6) using the assumed statistical properties of the piston error for the ensemble averaged RMS of the corrected wavefront we obtain:

$$\left\langle RMS_{fin}^2 \right\rangle = \frac{\sigma^2}{N} \sum_{j=1}^{N} \frac{1}{A_{dj}} \int \left[ \theta_j(\boldsymbol{\xi}) - IFs_j(\boldsymbol{\xi}) \right]^2 d^2\xi, \tag{7}$$

where $A_{dj}$ – aperture of the segment, $\boldsymbol{\xi} = \boldsymbol{x} - \boldsymbol{r}_j$. The residual RMS is proportional to the standard deviation of piston error. According to the Strong Low of Large Numbers the RMS before correction and the standard deviation of piston errors are related as[3]

$$\left\langle RMS^2 \right\rangle = \lim_{N \to \infty} RMS^2 = \sigma^2. \tag{8}$$

The ratio $\left\langle RMS_{fin}^2 \right\rangle / \left\langle RMS^2 \right\rangle$ does not depend on $\sigma$ and can serve as a quantitative measure of the RMS improvement by AO correction:

$$\gamma_{AO}^2 \equiv \frac{\left\langle RMS_{fin}^2 \right\rangle}{\left\langle RMS_{ini}^2 \right\rangle} = \frac{1}{N} \sum_{j=1}^{N} \frac{1}{A_{dj}} \int \left[ \theta_j(\boldsymbol{\xi}) - IFs_j(\boldsymbol{\xi}) \right]^2 d^2\xi = \frac{1}{N} \sum_{j=1}^{N} \gamma_{AOj}^2. \tag{9}$$

So far we did not use the assumption about the functions $IFs_j(\boldsymbol{x})$ or segments' shapes. If segments have identical shape and segment influence function doesn't depend on segment index, then the factor $\gamma_{AO}$ depends only on two parameters: number of actuators per segment side ($m$) and the shape of the influence function ($c$):

$$\gamma_{AO}(c, m) = \left[ 1 - 2 \frac{1}{A_d} \int \theta(\boldsymbol{\xi}) IFs(\boldsymbol{\xi}) d^2\xi + \frac{1}{A_d} \int IFs^2(\boldsymbol{\xi}) d^2\xi \right]^{1/2}. \tag{10}$$

For hexagonal all identical segments and the square DM actuators distribution there is no perfect matching between two geometries. The segment influence function is different for every segment. According to Eq.(9) to calculate $\gamma_{AO}$ one must averaged $\gamma_{AOj}^2$ over all existing actuators distribution over a hexagon. For calculations we averaged over four limiting cases, shown in Figure 4.

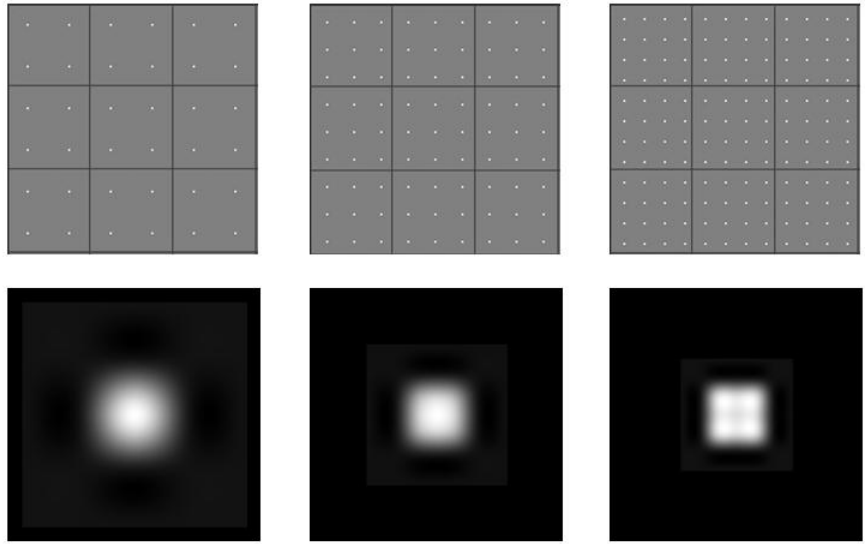

**Figure 2.** Dstribution of the actuators providing all identical segment influence functions (upper) and the corresponding segment influence function (bottom), *m*=1,2,3,4 (from left to right). Coupling factor *c* =0.

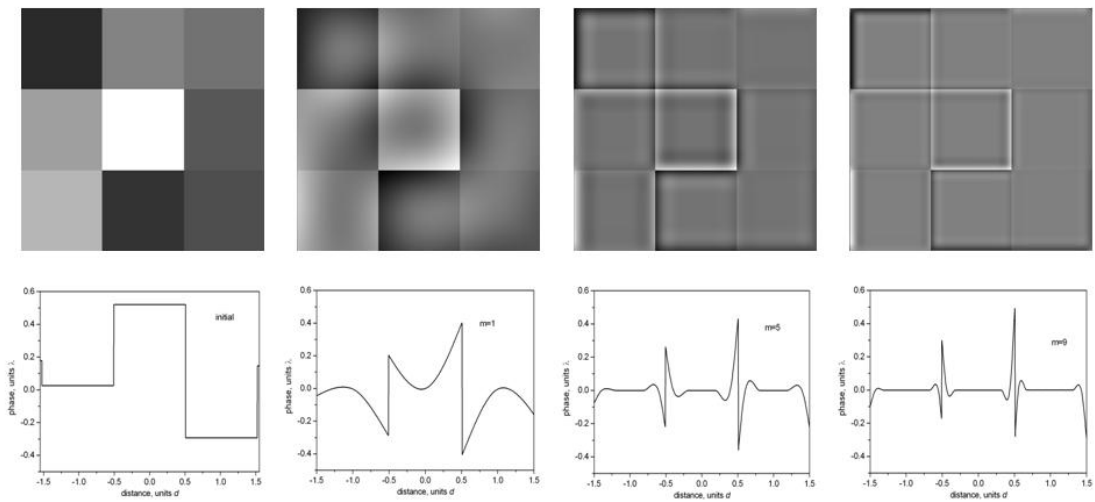

**Figure 3.** Initial and corrected phase in a gray scale (upper row) and the central vertical cut (low row). From left to right: initial phase, *m*=1,5,9. Coupling factor *c* =0.

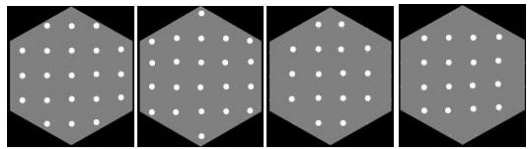

**Figure 4.** Four configurations of actuators used for calculation $\gamma_{AO}$ in case of hexagonal segments

The residual $\gamma_{AO}$ is shown in Figure 5 for square and hexagonal segments. In the latter case *d* is a flat to flat width of a hexagon, therefore parameter *m* is related to actuators' spacing as: $d/d_a = m\sqrt{3}/2$. Figure 5a shows $\gamma_{AO}$ as a function of a coupling coefficient. The optimum corresponds to zero coupling. The optimal point does not depend on *m*. Figure 5b shows $\gamma_{AO}$ as a function of $d/d_a$ for *c* =0. For the perfect matching of the segmentation and DM geometries (square segments in this example) the partial correction of segmentation errors by AO takes place already when the actuator separation equals to the size of a segment. If there are two or more actuators per segment length, the theoretical gain in RMS is about factor 2 for the large range of $c \geq 0$.

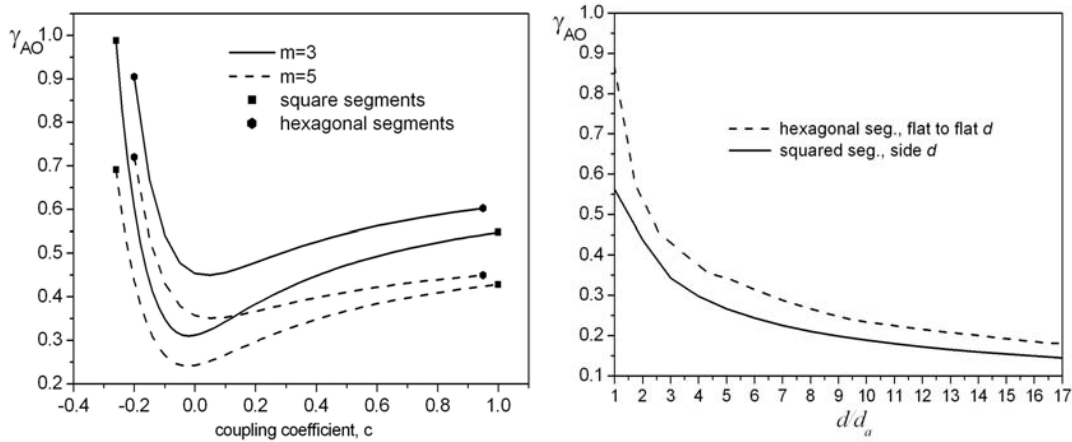

**Figure 5**. **Factor $\gamma_{AO}$, ratio between residual and final RMS for square and hexagonal segments as a function of actuators coupling coefficient (left) and number of actuator per segments side for c=0 (right).**

### 2.3 Residual power spectral density

The power spectral density (PSD) of the phase is statistically averaged modulus square of Fourier transform of the phase:

$$PSD(w) = \frac{1}{(A/\lambda)^2} \left\langle \left| \frac{1}{\lambda} \int \varphi(x) \exp\left(i\frac{2\pi}{\lambda} w \cdot x\right) d^2x \right|^2 \right\rangle. \tag{11}$$

It describes the distribution of the averaged power of the phase fluctuations over spatial frequencies. With respect to image formation the PSD is related to the blurring of the image, i.e. with the shape of an averaged point spread function (PSF). For the small errors the smooth component of the PSF (halo) can be approximated by the PSD itself. Therefore we relate the spatial frequency with the angular position in the image plane and calculate the PSD in the image plane coordinates (vector **w**). Accordingly we normalize the PSD by a factor equal to the intensity of the diffraction peak of the diffraction limited (non aberrated) PSF: $PSF_0(0) = (A/\lambda)^2$, where *A* is a pupil area and λ is a wavelength of an interest.

Substituting Eq. (6) into Eq. (11) for the PSD of the residual phase we obtain:

$$PSD(w) = \left\langle \left| \frac{1}{N} \sum_{j=1}^{N} \delta_j \exp\left(i\frac{2\pi}{\lambda} w \cdot r_j\right) \right|^2 \right\rangle \left| \frac{1}{A_d} \int [\theta(\xi) - IFs(\xi)] \exp\left(i\frac{2\pi}{\lambda} w \cdot \xi\right) d^2x \right|^2 \tag{12}$$

We assumed all identical $IFs_j(\xi)$ and we represented the telescope aperture as multiplication of the segment area $A_d$ by a number of segments. Performing calculations and using the adapted notations[4,5] we obtain the following final expression for the PSD:

$$PSD(w) = \frac{\sigma^2}{N} |t_a(w)|^2 \quad , \tag{13}$$

where $t_a(w)$ is Fourier transform of the function $[\theta(\xi) - IFs(\xi)]$. The *PSD* (halo of the *PSF*) is proportional to the variance of initial piston errors and inverse to the number of segments. Its shape is defined by the function $t_a(w)$. Taking into account the structure of the function $IFs(\xi)$ given by Eq.(5) and applying a shift theorem of Fourier transform, we can represent function $t_a(w)$ as:

$$t_a(w) = t(w) - if_a(w) \cdot gf_a(w), \tag{14}$$

where $t(w)$ is Fourier transform of segment pupil ($|t(w)|^2$ is PSF of one segment), $if_a(w)$ is Fourier transform of the influence function, and $gf_a(w)$ is a grid factor – Fourier transform of the actuators grid over the segment surface:

$$if_a(w) = \frac{m^2}{A_d} \int IF(\xi) \exp\left(i \frac{2\pi}{\lambda} w \cdot \xi\right) d^2\xi$$

$$gf_a(w) = m^{-2} \sum_{\substack{m \\ x_m \in \theta_j}} \exp\left(i \frac{2\pi}{\lambda} w \cdot \xi_m\right) \quad , \tag{15}$$

where $\xi_m = x - x_m$ and the factors $m^2$ and $m^{-2}$ are included to provide the unit value of both functions in a central point. In the case of square geometry shown in Figure 2 all function are factorizable into the *x* and *y* components, which allows us calculating the Fourier transform in one dimension only. The found expressions we summarize here:

$$t(w_{x(y)}) = \frac{2\sin(v/2)}{v}$$

$$gf_a(w_{x(y)}) = \frac{1}{m} \frac{\sin(v/2)}{\sin(v/2m)}, \tag{16}$$

$$if(w_{x(y)}) = \frac{4}{(2c+1)} \left[ (4c-1)\frac{\sin(v/m)}{(v/m)^3} + 6(2c-1)\frac{\cos(v/m)}{(v/m)^4} \right.$$

$$\left. - 3(3c-1.5)\frac{1}{(v/m)^4} - (2c-0.5)\frac{\sin(2v/m)}{(v/m)^3} - 3(c-0.5)\frac{\cos(2v/m)}{(v/m)^4} \right],$$

where $v = \frac{2\pi}{\lambda} w_{x(y)} d$. Function $|t_a(w_x,0)|^2$ is shown in Figure 6. According to Eq.(13) the conclusions concerning the efficiency of the correction can be done based on shape of $|t_a(w)|^2$. For comparison we also presented a case before the correction, when $t_a(w) = t(w)$. The area where DM partially corrects for the segmentation aberration is defined by the limits: $w_x \leq m\lambda/d$, $w_y \leq m\lambda/d$, which is below the AO cut off frequency $m\lambda/2d$ due to the optimal coupling (*c*=0). Outside this region $|t_a(w)|^2$ coincides with initial, non-corrected $|t(w)|^2$. The intensity of the halo within the corrected area (the first PSD peak) drops by a factor of 10 when *m* increases by a step of 5. The position of the main zeros, ($w_x = n\lambda/d$, $w_y = k\lambda/d$, $n, k = 1, 2...$) is not shifted by the correction. The two dimensional PSD is shown in a gray logarithmic scale in Figure 7.

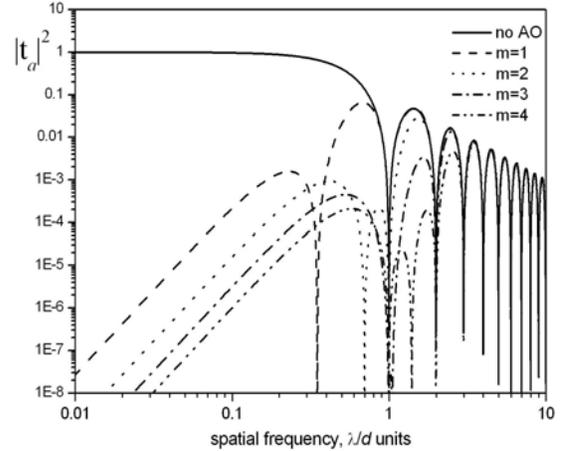

**Figure 6. Normalized PSD of corrected phase for different correction order in comparison with the initial phase PSD**

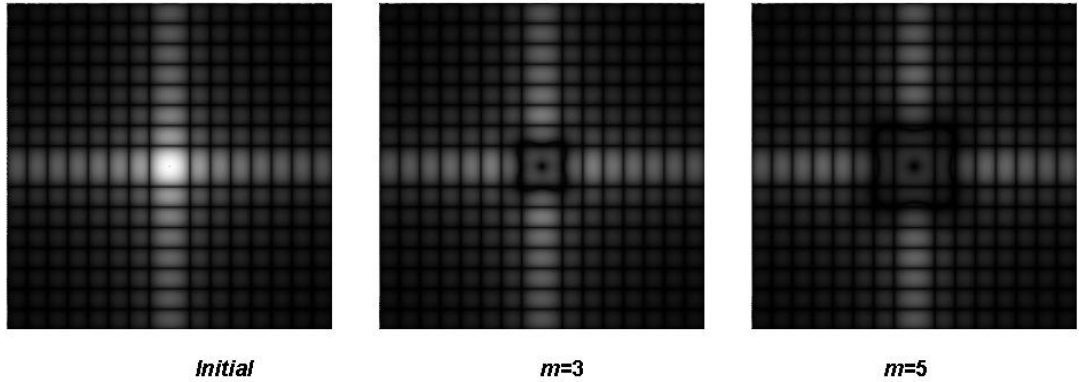

**Figure 7**. **PSD of the residual phase produced by segmented telescope when segmentation piston errors are corrected by deformable mirror; *m* is the number of actuators per segment side. The geometry of DM and segments is square.**

## 3. SIMULATIONS

### 3.1 Simulation model

The correction of piston phase errors with AO systems based on Shack-Hartmann sensors and Pyramid sensors has been simulated with an end-to-end model based on diffraction theory.[6] Only piston errors on square segments are considered, no atmosphere and no measurement noise are taken into account.
The general parameters used for the simulations are:
- Telescope diameter: $D = 8$ m.
- Actuators grid: 45×45.
- Influence functions: cubic spline, coupling: $c=0$
- Initial piston co-phasing errors: 30 nm RMS
- Segmentation: 15×15 square segments (segment size: $d=0.53$ m)
- Linear density of actuators $m=3$
- 2 actuators configurations (Figure 8)
    - **Configuration *a*:** No overlapping between actuators and segment edges (left)
    - **Configuration *b*:** The whole segment pattern is shifted by ½ actuator spacing in x and y so that some actuators centers coincide with segments edges (right)
- WFS configuration: 44×44, Fried geometry, sub-aperture size: $d_{sub}=0.18$ m.

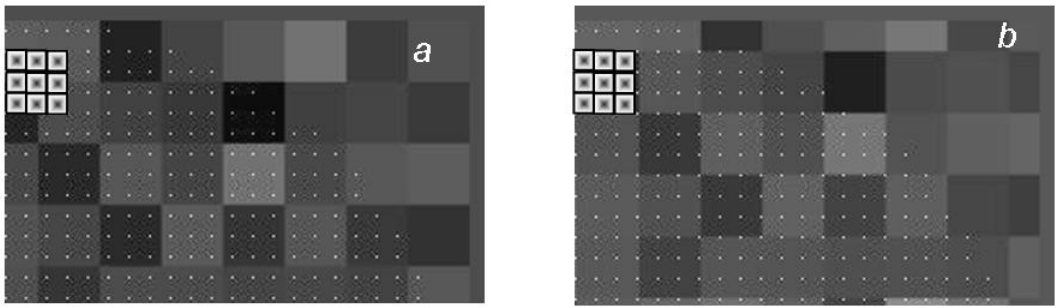

**Figure 8: Actuators grid with respect to segments used in the simulations. Configuration *a* (left): no actuators centers (points) on the segments edges. Configuration *b* (right): segments edges are covered by actuators centers. The Fried geometry of sub-apertures is shown in the upper left corner.**

In the Fried geometry (upper left corner of Figure 8*ab*) each WFS sub-aperture covers the region delimited by the position of the neighboring actuators' centers. Thus, in configuration *a*, each sub-aperture covers the wave-front over a discontinuity at the edges, whereas in configuration *b*, no discontinuity is covered by the sub-apertures area. A priori configuration *b* is the most unfavorable case: a piston-like phase errors would be completely invisible to a pure slope sensor. In configuration *a*, all discontinuities would be measured with the same highest accuracy by a slope sensor and would thus be ideal.

A real configuration is most probably an intermediate case of both configurations where some edges are well covered and others more poorly. However we use these two extreme cases to show the behavior of different types of WFS: a classical Shack-Hartmann, a spatially filtered Shack-Hartmann[7] (the diameter of the filter is $1.25\lambda/d_{sub}$) and a Pyramid sensor[8]. The control is in closed loop and the final results are the corrected wave-fronts at convergence in 10 iterations.

### 3.2 Results

First we compare the parameter $\gamma_{AO}$ obtained in simulations with actuators configuration *a* with the analytical model. The residual RMS in the case of a spatially filtered Shack-Hartmann WFS is 8.5nm, in the case of a Pyramid WFS 9.7nm. The RMS is reduced by a factor of 0.28 for Shack-Hartmann and by a factor of 0.32 for the Pyramid WFS. The theoretical value for $\gamma_{AO}(m=3, c=0) = 0.31$. The corrected wave front (part of the whole mirror) is shown in Figure 9 and a one-dimensional profile in Figure 10. Figure 11 represents a cut along x of the PSD of the corrected wave-front and Figure 12 shows the bi-dimensional PSD. The PSD is averaged over 10 realizations. The general behavior of the PSD, i.e. the reduction of phase errors in the low spatial frequency range is in agreement with the theory but with some variations depending on the wave-front sensor (see section 3.3). We can conclude here that the analytical model gives a sufficiently good estimation for RMS error improvement and for the level of the residual phase PSD.

The results presented in Figure 11 and Figure 12 show, however, that the WFS has a non negligible impact on the structure of PSD. Even though the RMS error in all cases is very similar, within one nanometer difference, the distribution of residual phase errors on the low-spatial frequency range ($< 1/2d_{sub}$) shows some structure that are significantly different from the theoretical one. This difference is particularly important for the eXtreme AO (XAO) since the PSD structure describes the residual halo one could obtain with a highly efficient coronagraph. In the next section, we show how the WFS can influence the structure of the PSD at these low spatial frequencies.

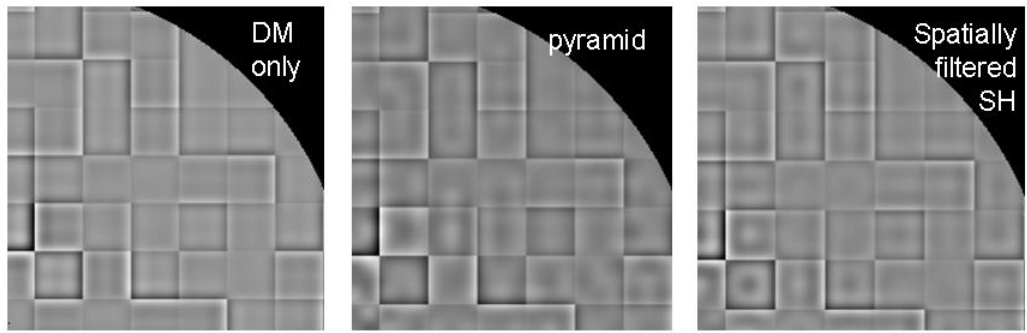

**Figure 9. Corrected wavefront according to the analytical model and obtained in a closed loop with two types of the wavefront sensor**

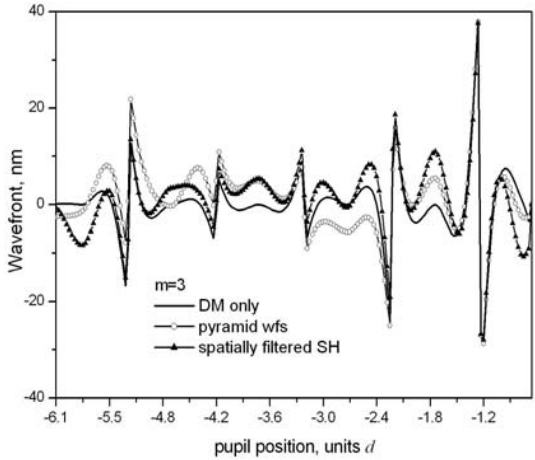
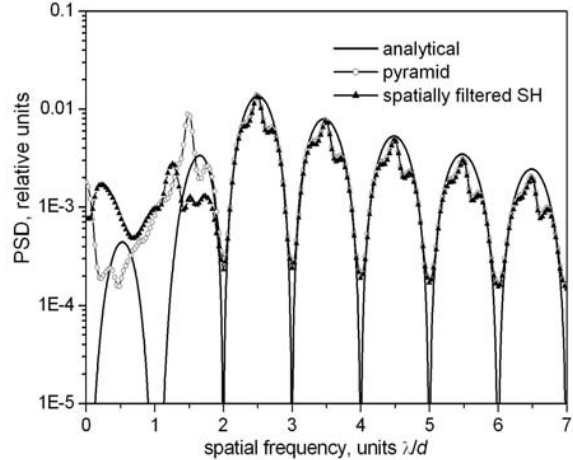

**Figure 10. Profile of the residual wavefront**

**Figure 11. PSD of the residual wavefront, normalized by $\sigma^2/N$ (similar to the Figure 6)**

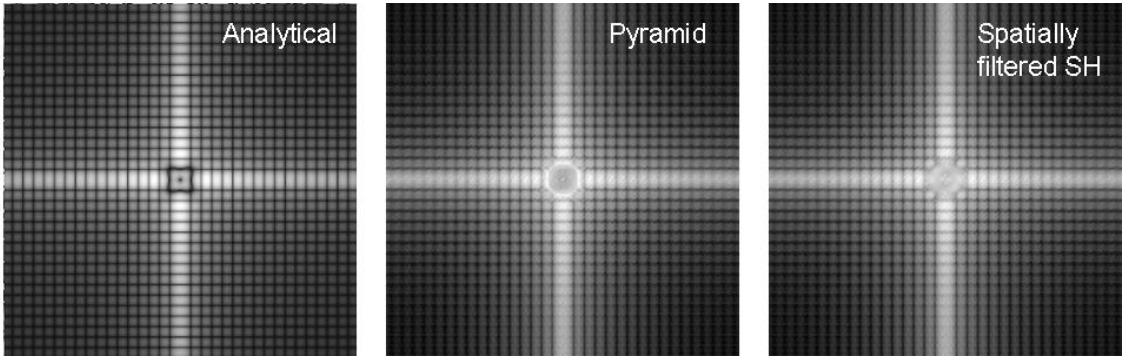

**Figure 12. PSD of the corrected wavefront according to the analytical model and obtained in a closed loop with two types of the wavefront sensor**

### 3.3 Wave-front sensor effect

In this section we study more in details the effect of the WFS on the PSD of the residual phase in the AO controlled region. For the configuration used in this simulation, the controlled spatial frequencies are $[0, 1/2d_{sub}=2.75\text{m}^{-1}]$. As already shown by several authors,[9,10] the PSD of the wave-front phase error in XAO, is a good estimation of the coronagraphic halo intensity, when phase errors dominate.

In order to show the effect on corrected coronagraphic PSFs we choose the imaging wave-length at 1.6 μm (H band) and represent thus the halo in normalized intensity (contrast) in function of angular separation with respect to the star position.

*3.3.1 Configuration a*

The results in terms of equivalent coronagraphic halo are presented in Figure 13 (bi-dimensional plot) and in Figure 14 (circularly averaged halo). The results with the classical Shack-Hartmann sensor are quite poor (Figure 13 and Figure 14, left). Even though the RMS error of the residual wave-front is only 1.5 nm worse for the Shack-Hartmann, the PSD and thus the intensity of the coronagraphic halo at angular separations less than 0.5 arcsec is more than 2 orders of magnitude higher. By comparing with a spatially filtered Shack-Hartmann (Figure 13 and Figure 14, right), the situation is completely different. The rms error is better than for the Pyramid sensor and the halo is much decreased in the central region controlled by AO. However, the halo with the spatially filtered Shack-Hartmann is still up to almost 10 times higher than the one with a Pyramid sensor for small angular separations (< 0.2 arcsec). Note also that the Shack-Hartmann performs better for larger angular separations (> 0.5 arcsec). This behavior is very similar to the noise propagation comparative characteristics of this sensors[11].

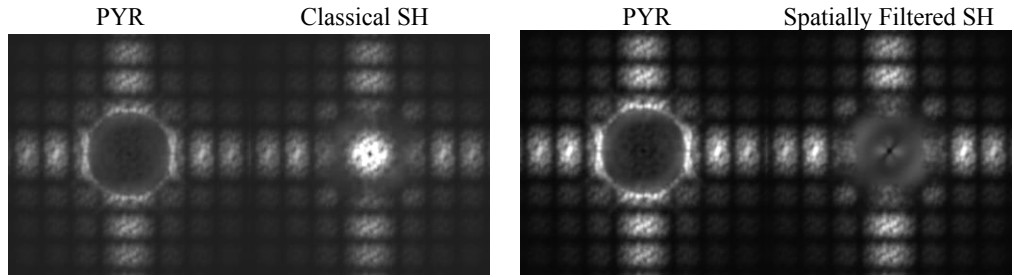

**Figure 13: Comparison of bi-dimensional PSD after the correction of piston errors for configuration *a*. Left: Pyramid (9.7 nm RMS) vs. classical Shack-Hartmann (11.2 nm RMS). Right: Pyramid vs. spatially filtered Shack-Hartmann (8.5 nm RMS). The Pyramid results are the same in the right and left picture, the difference is in gray scale for representation.**

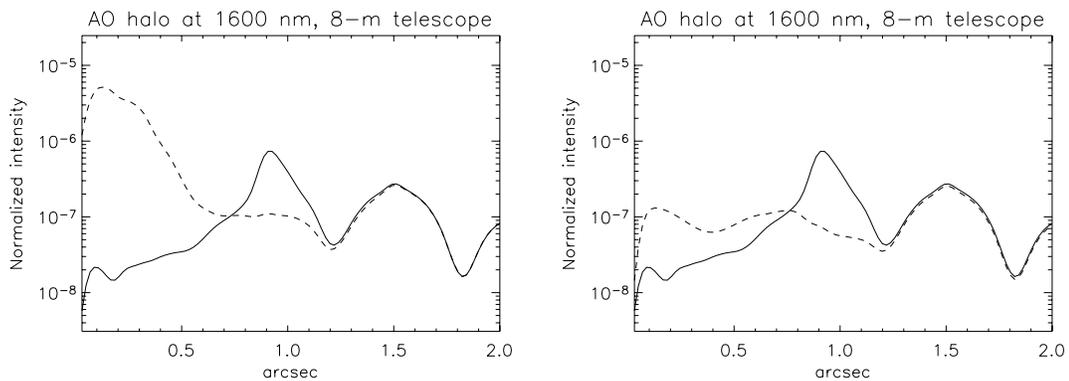

**Figure 14: Circularly averaged PSD from Figure 13. Left: Pyramid vs. classical Shack-Hartmann. Right: Pyramid vs. spatially filtered Shack-Hartmann. Solid line: Pyramid sensor. Dashed line: Shack-Hartmann.**

*3.3.2 Configuration b*

The general behavior is the same (Figure 15 and Figure 16). However, for the classical Shack-Hartmann the RMS errors is almost twice the one in configuration *a*. This is expected since configuration *b* is the worst for a slope sensor; the measurement is almost zero every where. Actually in a perfectly aligned system, no correction should occur. Here there is some correction because some mis-alignment due to numerical error exists in the simulation. For the Pyramid sensor, the halo rejection is almost as good as in configuration *a* with however some cross-like structure aligned with the Pyramid prism edges. The spatially filtered Shack-Hartmann permits to reject intensity in the center of the halo, but without reaching the performance of a Pyramid sensor.

The difference between the Pyramid sensor and the classical Shack-Hartmann sensor can be explained by the fact that the Pyramid sensor is a 'phase-like' sensor,[12] more exactly, it measures the Hilbert transform of the phase. The measurements are thus the result of a convolution product, so that a piston phase step always produces a nonzero measurement on the Pyramid detector.

In the spatially filtered Shack-Hartmann the effect of the spatial filter is to smooth out the steps in the incoming phase. The slope of the phase is thus nonzero inside a segment area (Figure 17) unlike in a classical Shack Hartmann method. That signal is sufficient to close the loop.

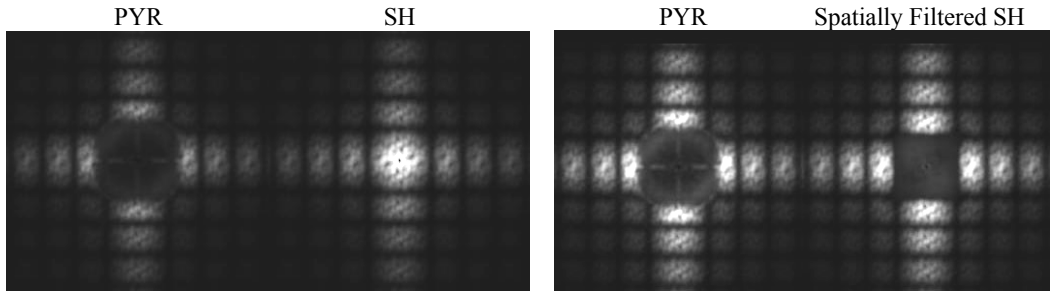

**Figure 15:** Comparison of bi-dimensional PSD after correction of piston phase errors for configuration b. Left: Pyramid (11.4 nm RMS) vs. classical Shack-Hartmann (18.2 nm RMS). Right: Pyramid vs. spatially filtered Shack-Hartmann (11.3 nm RMS). The Pyramid results are the same in the right and left picture (difference gray scale for representation).

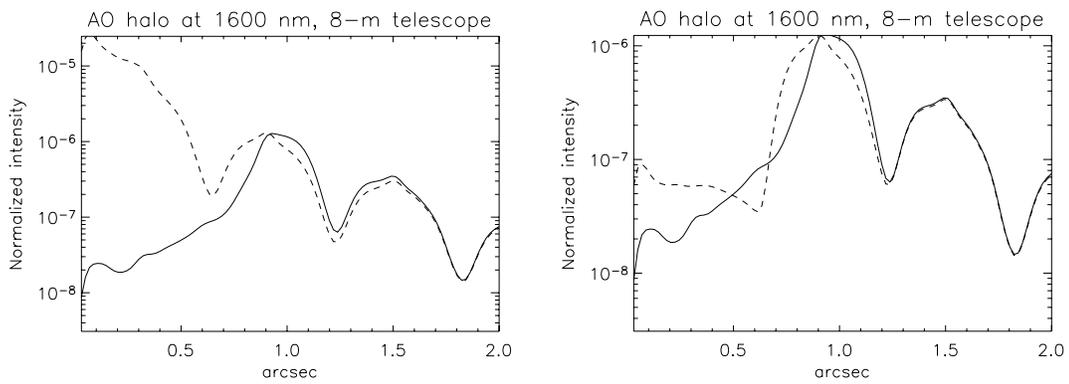

**Figure 16:** Comparison of circularly averaged PSD after correction of piston phase rrors for configuration b. Left: Pyramid vs classical Shack-Hartmann sensor. Right: Pyramid vs spatially filtered Shack-Hartmann sensor. Solid line: Pyramid sensor. Dashed line: Shack_Hartmann.

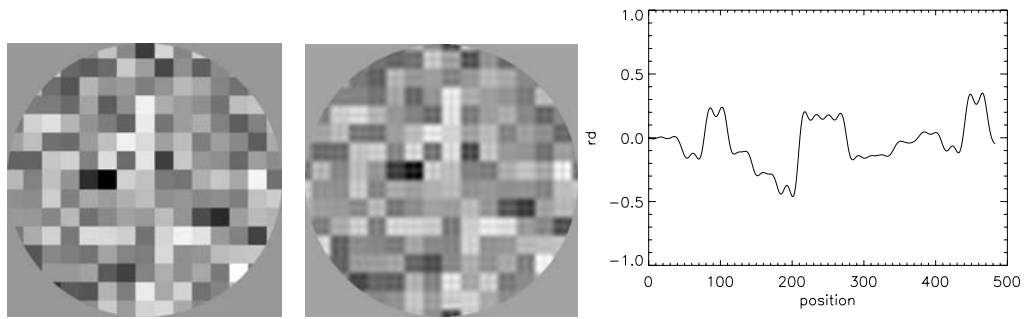

**Figure 17:** Initial wave-front reaching the Shack-Hartmann wave-front sensor. Left: classical Shack-Hartmann. Center: Spatially filtered Shack-Hartmann (1.25$\lambda/d_{sub}$ filter diameter). Right: cut in spatially filtered wave-front.

## 4. CONCLUSION

The ability of the AO system to correct for the segmentation errors strictly depends on the design of the deformable mirror and the wavefront sensor. The density of the actuators, the coupling factor and the geometry of actuators distribution are essential factors influencing the performance. In a case of a perfect matching between the geometry of the DM and the segmentation and the optimal (zero) coupling, the correction takes place already with one actuator per segment ($m$=1). For the typical ratio $m$=6 (segment 1.5 m, actuator pitch 25 cm) the RMS is reduced in 3 times for hexagonal segments and in 4 times for the square segments, both in case of a square actuators distribution. The study of the residual phase PSD demonstrates the correction within the limits set by a cut-off frequency of the deformable mirror, as expected. Within this region the level of the PSD, related to the PSF halo, decreases by a factor of 10 when the linear density of actuators is increased by 5.

Although the analytical investigation did not concern the wavefront sensing aspects, it shows a good agreement with the numerical results. The simulations have also investigated the importance of the type of WFS. A classical Shack-Hartmann that measures the slope of the wave-front performs quite poorly to correct for piston-like co-phasing errors since the measurements are almost zero everywhere except on the edges. Conversely, a spatially filtered Shack-Hartmann permits to get an error signal that is nonzero everywhere and thus to close the loop on piston errors with much better performance. The pyramid sensor, which is known to have similar properties to a phase sensor, shows the best performance for the correction at small angular separations. The Shack-Hartmann however performs better at high angular separations.

The next steps in the study will consist in performing end-to-end simulations for a larger diameter and with higher actuators per segments density. Atmospheric turbulence and measurement noise will also be included to test the real sensitivity of the different sensors.

## ACKNOWLEDGMENTS

The presented study is accomplished in the Framework Programme 6, ELT Design Study, contract No 011863.